\documentclass[a4paper,11pt]{article}
\usepackage{pos}
\usepackage{placeins}
\usepackage[capitalise]{cleveref}
\usepackage{slashed}

\def\<#1>{\mathinner{\langle#1\rangle}}
\newcommand{\Nf}{N_f}
\newcommand{\Det}{\mathrm{Det}}
\newcommand{\Tr}{\mathrm{Tr}}
\newcommand{\iu}{\mathrm{i}}
\newcommand{\dr}{\mathrm{d}}
\newcommand{\phib}{\phi^{(\mathrm{b})}}
\newcommand{\phif}{\phi^{(\mathrm{f})}}
\newcommand{\phix}{\phi^{(\mathrm{x})}}
\newcommand{\trqinvn}[1]{\Tr \,Q^{-#1}}
\newcommand{\expConstrained}[2][{\phix}]{\bigl< #2 \bigr>_{#1}}

\newcommand{\expcc}[1][{x}]{\bigl< \bar \psi \psi  \bigr>_{\phi^{(\mathrm{#1})}}}
\newcommand{\constraintPotential}[1][x]{\Omega_{\mathrm{#1}}}
\newcommand{\trone}{\mathcal{M}}
\newcommand{\trtwo}{\chi}
\newcommand{\trthree}{\gamma}

\usepackage[acronym]{glossaries-extra}
\glssetcategoryattribute{acronym}{nohyperfirst}{true}
\setabbreviationstyle[acronym]{long-short}

\newacronym{gn}{GN}{Gross-Neveu}
\newacronym{qcd}{QCD}{Quantum chromodynamics}
\newacronym{hs}{HS}{Hubbard-Stratonovich}

\title{Flattening of the quantum effective potential in fermionic theories}

\author[a]{Gergely Endrődi}
\author[b,c]{Tamás G. Kovács}
\author[a]{Gergely Markó}
\author*[a]{Laurin Pannullo}
\affiliation[a]{Fakultät für Physik, Universität Bielefeld,
  D-33615 Bielefeld, Germany}

\affiliation[b]{Department of Theoretical Physics, Eötvös Loránd University,
Pázmány Péter sétány 1/A, H-1117, Budapest Hungary}

\affiliation[c]{Institute for Nuclear Research,
	 Bem tér 18/c, H-4026 Debrecen, Hungary}

\emailAdd{lpannullo@physik.uni-bielefeld.de}

\abstract{We present methods to constrain fermionic condensates on the level of the path integral, which grant
	access to the quantum effective potential in the infinite volume limit.
	In the case of a spontaneously broken symmetry, this potential possesses a manifestly flat region, which is inaccessible to the standard
	approach on the lattice. However, by constraining the appropriate order parameters such as the chiral condensate, one is then able to probe the flat region. We demonstrate our
	method of constraining fermionic condensates in the 2-dimensional Gross-Neveu model, which exhibits a spontaneously broken chiral symmetry. We show how the potential flattens for increasing volume and that
	the flat region is dominated by inhomogeneous field configurations.
	}

\FullConference{The 40th International Symposium on Lattice Field Theory (Lattice 2023)\\
July 31st - August 4th, 2023\\
Fermi National Accelerator Laboratory\\}

\begin{document}
\maketitle

\section{Introduction}
	In this work, we consider fermionic theories that spontaneously break chiral symmetry, resulting in a non-zero chiral condensate.
	The standard procedure to investigate the breaking of such a symmetry on the lattice would be to introduce an explicit breaking of the symmetry and then taking the double limit of infinite volume and subsequently zero explicit breaking. 
	Such a procedure can be extremely expensive in theories like \gls{qcd}.
	An alternative approach would be to investigate the quantum effective potential $\Gamma$, which is the Legendre transform of the free energy with respect to the expectation value of the order parameter -- in our case the chiral condensate. 
	It then incorporates all quantum effects of the theory.
	This manifestly convex potential features a flat region in the case of a broken symmetry, with the edges of the flat region corresponding to the ground state manifold.
	Thus, if one has access to this object, it is rather trivial to investigate whether the symmetry is broken and what the value of the order parameter is.

	While one cannot access this quantity directly in lattice simulations, it is possible to employ the constraint path integral formalism.
	Here, one introduces a constraint  of the form $\delta(\phi - \varphi)$ on a given order parameter $\varphi$ to the path integral.
	The resulting constraint path integral $\mathcal{Z}_\phi$ only integrates over the configurations that fulfill the constraint, i.e., $\varphi=\phi$.  
	The so-called constraint potential $\Omega(\phi)= - \ln \mathcal{Z}_\phi/V$ can be shown to coincide with the quantum effective potential $\Gamma(\phi)$ in the infinite volume limit \cite{ORaifeartaigh:1986axd}.
	This gives an alternative approach to investigate a spontaneous symmetry breaking, where one only has to take the infinite volume limit instead of the double limit that we discussed before.
	While a constraint of bosonic order parameters has been successfully applied on the lattice \cite{Fodor:2007fn,Endrodi:2021kur}, it has not been used so far for fermionic order parameters such as the chiral condensate.
	In this work, we present methods to constrain the chiral condensate and obtain the constraint potential.
	We test these methods in the $2$-dimensional \gls{gn} model \cite{Gross:1974jv} in the $\Nf\to\infty$ limit.
	While this presents a rather simple setup, our method is applicable to other fermionic theories such as \gls{qcd}.
	This work presents preliminary investigations of our developments that we will discuss in full detail in an upcoming publication \cite{Endrodi:2023XXX}.

\section{The Gross-Neveu model}

	The $2$-dimensional \gls{gn} model \cite{Gross:1974jv} is a so-called four-Fermi theory with its bosonized Euclidean action obtained from a \gls{hs} transformation given by 
	\begin{align}
		S_\sigma = \int \mathrm{d}^2x \left[- \bar \psi(x) \left(\slashed{\partial}+\sigma(x)\right) \psi(x) + \tfrac{\Nf}{2g^2}  \sigma^2(x)\right].
	\end{align}
	where $\bar \psi, \psi$ are fermion spinors with $N_\gamma=2$ components and $\Nf$ species, and the coupling $g^2$ controls the coupling strength of the scalar four-Fermi interaction $ \left(\bar \psi \psi \right)^2$ in the original action before bosonization.
	The bosonic field $\sigma$ is an auxiliary field that is introduced by the \gls{hs} transformation and its expectation value is proportional to the chiral condensate as given by the Ward identity
	\begin{align}
		\langle\bar \psi \psi  (x) \rangle = \tfrac{\Nf}{g^2}\langle \sigma(x) \rangle. \label{eq:Ward}
	\end{align}
	One can integrate over the fermionic degrees of freedom in the path integral to obtain 
	\begin{align}
		\mathcal{Z}=\int \mathcal{D} [\sigma] \, \mathrm{e}^{-S_\mathrm{eff}}, \quad  S_\mathrm{eff}=\int \mathrm{d}^2x\,  \tfrac{N_f}{2g^2}  \sigma^2(x) - \ln \Det\, Q[\sigma], \quad 
		Q[\sigma]=\slashed{\partial} + \sigma.
	\end{align}
	
	An often considered limit is the limit $\Nf \to \infty$, which suppresses all quantum fluctuations for the bosonic field $\sigma$.
	In this limit, the only configurations $\Sigma$ that contribute to the path integral are the ones with the largest weight. 
	These are the ones that minimize the effective action $S_\mathrm{eff}$.
	Thus, the calculation of the path integral reduces to a minimization problem.
	We exclusively consider this limit in this work.
	
	The theory exhibits a discrete chiral symmetry under which the fermionic and bosonic fields transform as 
	\begin{align}
		\psi \to \gamma_5 \psi,\quad \bar \psi  \to -\bar \psi \gamma_5, \quad \sigma \to - \sigma.
	\end{align}
	This symmetry is spontaneously broken in the $\Nf \to \infty$ limit and one finds a non-zero expectation value $\langle \bar \psi \psi \rangle$.
	We renormalize the theory, by requiring that the expectation value of $\sigma$ assumes a homogeneous non-zero value, i.e., $\langle \sigma \rangle = \sigma_0$, which is achieved by choosing an appropriate value of the coupling (see Refs.~\cite{Koenigstein:2021llr} for further details on the renormalization).
	
	In this work, we consider a naive lattice discretization of the \gls{gn} model.
	We refer to Ref.~\cite{Lenz:2020bxk} for a detailed discussion of the \gls{gn} model within this discretization.
	
\section{Constraining the Gross-Neveu model}

	We are interested in constraining the expectation value of the chiral condensate and obtaining the effective potential as a function of this expectation value.
	We consider two options to achieve this:
	\begin{enumerate}
		\item a \textit{fermionic constraint}, where we constrain the chiral condensate directly,
		\item a \textit{bosonic constraint}, where we constrain the auxiliary field, which is linked to the chiral condensate via the Ward identity \labelcref{eq:Ward}.
	\end{enumerate}
	\subsection{Fermionic constraint}
	
	With the fermionic constraint, we modify the bosonized path integral by introducing a constraint on the spatial average of the condensate to obtain
	\begin{align}
		\mathcal{Z}_{\phif}=\int \mathcal{D}[\bar\psi,\psi,\sigma] \ \mathrm{e}^{-S_\sigma}\, \delta\!\left( \tfrac{1}{V} \int \mathrm{d}^2x\, \bar \psi(x) \psi(x) - \phif N_f \right). \label{eq:fermionic_constraint_Z}
	\end{align}
	This type of constraint involves a combination of Grassmann numbers in a Dirac delta distribution.
	A priori, it is not clear whether this is a well-defined object.
	However, if we choose a representation of the Dirac delta as
	$
		\delta(x)= \int_{-\infty}^{\infty} \tfrac{\dr\eta}{2 \uppi} \, \mathrm{e}^{\iu\eta x},
	$
	one recognizes that it merely requires that arbitrary powers of $x$ are non-vanishing.
	This is realized if $\bar \psi$ and $\psi$ have infinitely many degrees of freedom as, e.g., in the infinite volume, and thus arbitrary powers of $\tfrac{1}{V}\int \dr x \, \bar \psi (x) \psi (x)$ do not vanish.
	This aspect will be discussed in greater detail in Ref.~\cite{Endrodi:2023XXX}.

	Applying this expansion to the path integral with the fermionic constraint and a subsequent integration over the fermionic degrees of freedom results in the reduced path integral
	\begin{align}
		\mathcal{Z}_{\phif}= 	\int\!\! \mathcal{D}\sigma  \int_{-\infty}^{\infty} \tfrac{\dr\! \eta}{2 \uppi}  \ \mathrm{e}^{-S'_\mathrm{eff}(\eta)[\sigma]-\iu \eta \phif N_f}\, , \quad 
		S'_\mathrm{eff}(\eta)[\sigma] = -\ln \Det \left(Q\big[\sigma\big]+\tfrac{ \iu \eta}{V} \right)   + \int \mathrm{d}^2 x \, \tfrac{\Nf \sigma^2}{2g^2}.
	\end{align}
	One cannot perform the $\eta$-integral analytically in this form. 
	We can, however, expand the determinant in powers of $1/V$ to obtain
	\begin{align}
		S'_\mathrm{eff}(\eta)[\sigma] = -\ln \Det \left(Q[\sigma] \right) - \tfrac{\iu \eta}{V}\trqinvn{1}  - \tfrac{\eta^2}{2V^2}\trqinvn{2} + \mathcal{O}\!\left(\tfrac{\eta^3}{V^3}\right)  +  \int \mathrm{d}^2 x \, \tfrac{\Nf \sigma^2}{2g^2}.
	\end{align}
	Neglecting terms of order $\mathcal{O}\left(\tfrac{1}{V^3}\right)$, allows us to carry out the $\eta$-integral in a closed form to obtain the constraint path integral 
	\begin{align}
		\mathcal{Z}_{\phif} =  \mathcal{N}	\int\!\! \mathcal{D}\sigma \,  \mathrm{e}^{-S_\mathrm{eff}[\sigma]-\tfrac{\Nf \left(\phif  - \trone\right)^2}{2\trtwo}- \tfrac{1}{2} \ln \trqinvn{2} } , \quad \trone \equiv \tfrac{1}{VN_f} \trqinvn{1} ,\quad \trtwo \equiv -\tfrac{1}{VN_f} \trqinvn{2}, \label{eq:fermionic_constraint_path_integral}
	\end{align}
	where $\mathcal{N}$ is an irrelevant normalization constant and the additional $\ln \trqinvn{2}$ term is suppressed in the limit $\Nf \to \infty$.
	The relevant configurations $\Sigma$ in this limit are the ones that minimize the complete exponent of the weight instead of only $S_\mathrm{eff}$. 
	One can intuitively understand that the modifications to the weight of the path integral in \cref{eq:fermionic_constraint_path_integral} punish configurations, where $\trone$ deviates from $\phif$.
	As the expectation value of $\trone$ would be the chiral condensate in the unconstrained case, this result appears plausible.
	
	In the presence of the constraint, it is, however, important to distinguish ${\cal M}$ from the chiral condensate, which reads then
	\begin{align}
		\frac{\expConstrained[\phif]{\bar \psi \psi }}{V N_f} = \phif + \bigg\langle \frac{(\phif-\trone)^2}{\trtwo} \frac{\trthree}{\trtwo}\bigg\rangle_{\phif} - \frac{1 }{V N_f}\bigg\langle \frac{\trthree}{\trtwo}\bigg\rangle_{\phif} , \quad \trthree \equiv \tfrac{1}{V N_f} \trqinvn{3} \label{eq:cc_fermionic_constraint}.
	\end{align}
	This might be surprising, since from \cref{eq:fermionic_constraint_Z}, one would have expected the chiral condensate to be $\phif$.
	The volume expansion destroys this exactness and one retains additional terms as modifications, which stem from the exchange of the infinite volume limit and the $\eta$-integration.
	However, we will see in \cref{sec:results} that these corrections are actually small in the relevant regions of $\phif$.
	Moreover, one finds that the Ward identity \labelcref{eq:Ward} is preserved even after the volume expansion.

\subsection{Bosonic constraint}

	The bosonic constraint, introduces a constraint on the spatial average of the field $\sigma$ to the path integral as
	\begin{align}
		\mathcal{Z}_{{\phib}}=\int \mathcal{D}\big[\bar\psi,\psi,\sigma\big] \ \mathrm{e}^{-S_\sigma}\, \delta\!\left( \tfrac{1}{V} \int \mathrm{d}x\, \sigma(x) - \phib \right).
	\end{align}
	Even though, we do not constrain the chiral condensate directly, one might expect that it is constrained nevertheless due to the Ward identity \labelcref{eq:Ward}.
	However, one finds that this Ward identity receives an extra contribution in the constrained path integral with the modified relation given by
	\begin{align}
		\frac{\langle\bar \psi \psi  \rangle_{\phib}}{V} = \frac{\Nf}{g^2}\phib - \frac{\partial \constraintPotential[b]\left(\phib\right)}{\partial \phib} \equiv  \frac{\Nf}{g^2}\phib - j_\mathrm{b}\left(\phib\right), \label{eq:Ward_bosonic}
	\end{align}
	with $\constraintPotential[b] = - \ln \mathcal{Z}_{\phib}$
	Thus, we expect that this bosonic constraint constrains the chiral condensate up to $j_\mathrm{b}$, which is the derivative of the constraint potential with respect to the constraint parameter.
	Moreover, in contrast to the fermionic constraint, one finds that $\langle\bar \psi \psi  \rangle_{\phib}=V \expConstrained[\phib]{\trone}$ without any corrections.
	Bosonic constraints have been considered in various studies \cite{Fodor:2007fn,Endrodi:2021kur} and their application has been well understood.
	We exploit this type of constraint in the special setup of the \gls{gn} model to benchmark the ability of the fermionic constraint to constrain the chiral condensate, which we focus on in this work.
	As the bosonic constraint works on the level of the bosonic fields, which are the degrees of freedom in which the minimization of the action is done, we can enforce it numerically in the minimization routine.
	The same is true for a Monte-Carlo simulation, where the $\sigma$-field would be the degree of freedom that is integrated over.

\section{Results of the constrained Gross-Neveu model}
\label{sec:results}

	As a first test of the constraining methods, we consider a single coupling $g^2\approx0.49203$ on two lattices $V=L^2\in\{40^2,60^2\}$.
	For this coupling, the bosonic vacuum expectation value with the unconstrained path integral is $\langle \sigma \rangle=\sigma_0=0.5=1.01620 \, g^2$. 
	In order to facilitate the numerical minimizations, we restrict the $\sigma$-field to be homogeneous in one of the two directions.
	
\begin{figure}[tb]
		\centering
		\includegraphics[width=\linewidth]{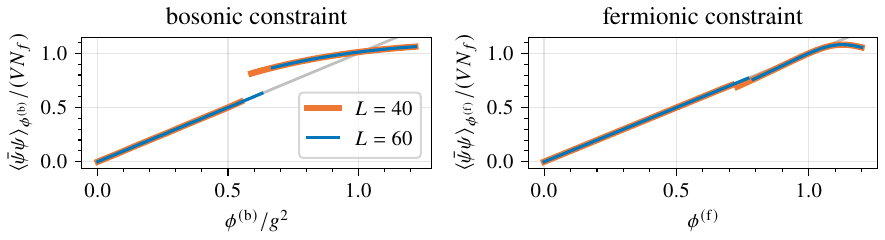}
		\caption{The expectation value $\expcc$ for the two constraint methods.}
		\label{fig:gnlargennaivechiralcondensate}
	\end{figure}

	We start our investigation by calculating $\expcc[x]$ for both constraints as shown in \cref{fig:gnlargennaivechiralcondensate} to estimate the extent of the consistency of the constraints.
	With the bosonic constraint, we find that the chiral condensate is given by $\phib/g^2$ up to a  maximum value in $\phib$, where the additional term in \cref{eq:Ward_bosonic} contributes, i.e., $j\neq0$.
	The chiral condensate being constrained that strongly in some range of $\phib$ even though we constrained the bosonic field, is certainly remarkable.
	One finds that this maximum value of $\phib$ increases with increasing volume.
	We find a similar situation with the fermionic constraint, where the chiral condensate is constrained again up to a maximum value in $\phif$, which increases also with volume.
	However, the deviation from $\phif$ is minimal beyond this point and the chiral condensate is still constrained approximately.
	This shows that the chiral condensate is controlled considerably better by the fermionic constraint compared to the bosonic constraint.
	For even larger $\phif$ (beyond $\sigma_0/g^2$), one finds that there is a large deviation of $\expcc[f]$ from $\phif$, which appears to be independent of the volume.
	This signals that the additional terms in \cref{eq:cc_fermionic_constraint} become relevant.
	
\begin{figure}[tb]
		\centering
		\includegraphics{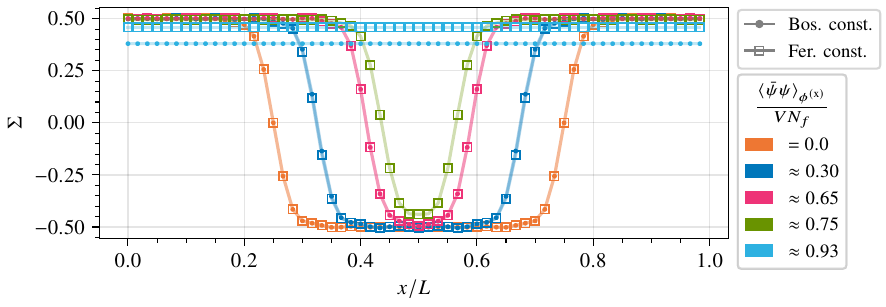}
		\caption{The minimizing field configuration $\Sigma$ for several values of $\expcc$ with the bosonic and fermionic constraint for $L=60$.}
		\label{fig:gnlargennaivecomparefieldconfigccscan}
	\end{figure}

	To understand how the constraints works and why the jumps at the intermediate values of $\phix$ happen, we consider the field configurations $\Sigma$ that we find in the numerical minimization for $L=60$ in \cref{fig:gnlargennaivecomparefieldconfigccscan}.
	We show the configurations for both constraint types each at similar values of $\expcc[x]$.
	For small values of $\expcc$, we find inhomogeneous configurations in both cases, which interpolate between the minima of the classical double well potential at $ \pm \sigma_0 = \pm 0.5$.
	The bosonic field $\Sigma$ mostly stays in one of the two minima and transitions through the unfavored regions with a tunneling profile that is independent of $\expcc$.
	For increasing value of $\expcc$, the ratio  of space that the bosonic field spends in one of the minima changes to accommodate the change in the constraint.
	At a certain value of $\expcc$, the required ratio cannot be fulfilled anymore leading to the observed breakdown of the constraint. 
	The approximate nature of the fermionic constraint enables configurations such as the green curve, where the field does not tunnel all the way.
	These configurations correspond to a value of the chiral condensate just before the jump at $\phif\approx0.75$ of the $L=60$ curve in the right plot of \cref{fig:gnlargennaivechiralcondensate}.
	For large $\expcc$, one finds homogeneous field configurations for both constraining methods.
	We find that the tunneling profile has a fixed physical width and, thus, for larger volumes more extreme ratios of space in the two minima can be fulfilled.
	
\begin{figure}[tb]
		\centering
		\includegraphics[width=\linewidth]{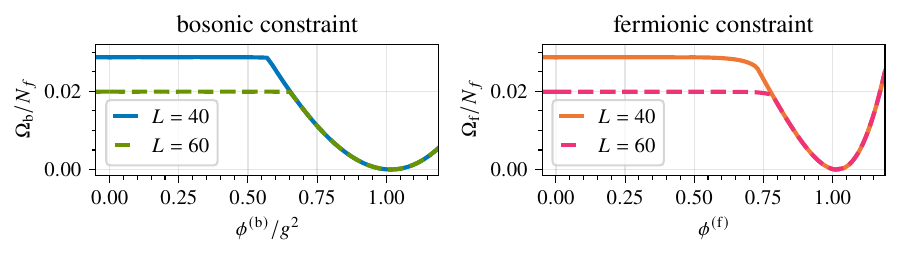}
		\caption{The constraint potential $\constraintPotential[x]$ as a function of the constraint parameter with the bosonic and fermionic constraint for $L\in\{40,60\}$. }
		\label{fig:gnlargennaiveomega}
	\end{figure}

	Now we turn to the constraint potential $\constraintPotential[x] = -\ln \mathcal{Z}_{\phix}/V$. 
	In a full Monte-Carlo simulation, one would measure $j_x$ -- the derivative of $\constraintPotential[x]$ w.r.t.~to the constraint parameter -- and reconstruct $\constraintPotential[x]$ from this \cite{Endrodi:2021kur}.
	In the considered $\Nf \to \infty$ limit, however, we can also directly calculate $\constraintPotential[x]$, because there is only a single configuration that is contained in the path integral.
	The two methods yield the same result within numerical accuracy.
	\cref{fig:gnlargennaiveomega} depicts the constraint potential for both constraining methods as a function of the constraint parameter.
	Both methods feature a flat region for small constraint parameters, which extends to larger constraint parameters for larger volumes.
	This regime features the inhomogeneous field configurations shown in \cref{fig:gnlargennaivecomparefieldconfigccscan}.
	The value of the flat region is finite but approaches the value corresponding to the minimum for increasing volume.
	The flat region will extend all the way to the minimum in the infinite volume limit.
	This is the behavior that we expect from the quantum effective potential, which agrees with the constraint potential for $V\to\infty$.
	We notice that the deviation from the flat behavior is non-analytic for the bosonic constraint and smooth for the fermionic constraint, which reflects the approximate nature of this constraint.
	\begin{figure}[tb]
		\centering
		\includegraphics[width=\linewidth]{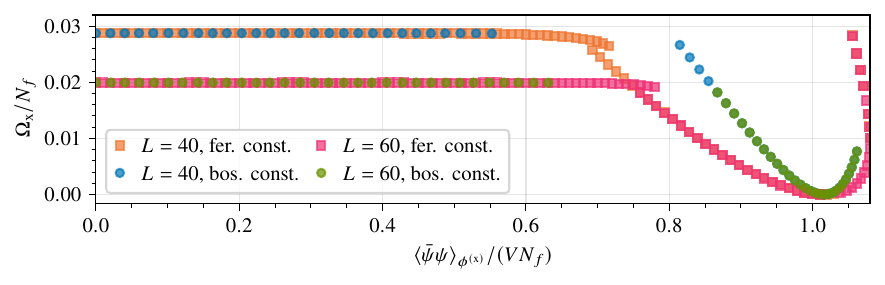}
		\caption{The constraint potential $\constraintPotential[x]$ as a function of $\expcc$ with both constraint types for $L\in\{40,60\}$. }
		\label{fig:gnlargennaiveomegacc}
	\end{figure}
	Next, we consider the constraint potential as a function of $\expcc$ instead of the constraint parameter, as this enables a real comparison of the two methods.
	This is done in \cref{fig:gnlargennaiveomegacc}, where the results in the flat region from different constraint methods for the same volume agree  within numerical accuracy.
	For the fermionic constraint, one finds that $\constraintPotential[f]$ is double-valued for some values of $\expcc[f]$. 
	There is, however, no stability issue as this is only due to the fact that two values of $\phif$ can correspond to a single value of $\expcc[f]$ in the region where it jumps in \cref{fig:gnlargennaivechiralcondensate}.
	We conclude that the fermionic constraint is actually able to constrain the chiral condensate in order to calculate the constraint potential. 
	As expected, this flattens for increasing volume and we expect it to be equivalent to the quantum effective potential in the infinite volume limit.

\section{Conclusion}

	In this work, we presented two methods to constrain the chiral condensate on the level of the path integral and used these to successfully calculate the constraint potential as a function of the expectation value $\langle \bar{\psi} \psi \rangle$.
	While the bosonic constraint was used in this form in others contexts, the fermionic constraint is an entirely new approach, which directly constrains the fermion bilinear $\bar{\psi}\psi$.
	Neither of the methods are limited to the $\Nf \to \infty$ limit in which we tested the methods, but could also be used in Monte-Carlo simulations. 
	It was shown that such constrained simulations are an excellent tool to investigate spontaneous symmetry breaking on the lattice with a lesser numerical effort compared to the traditional method with an explicit breaking \cite{Endrodi:2021kur}.
	Moreover, the fermionic constraint is not unique to the \gls{gn} model and can be implemented similarly in other fermionic theories. 
	This includes \gls{qcd}, where it could provide a valuable alternative to the standard approach in full lattice simulations.
	This discussion and results of the constrained chiral \gls{gn} model featuring a continuous chiral symmetry will be presented in Ref.~\cite{Endrodi:2023XXX}.

\FloatBarrier
\section*{Acknowledgments}

	We acknowledge useful discussions with Bastian Brandt.
	We acknowledge the contribution of Marc Winstel to the code framework that was used for the numerical minimization of the constrained Gross-Neveu model.
	G.E., G.M.~and L.P.~acknowledge the support of the Deutsche Forschungsgemeinschaft (DFG, German Research Foundation) through the collaborative research center trans-regio CRC-TR 211 “Strong-interaction matter under extreme conditions”– project number 315477589 – TRR 211.
	L.P.~acknowledges the support of the Helmholtz Graduate School for Hadron and Ion Research and of the Giersch Foundation.

\bibliographystyle{JHEP}
\bibliography{bib}

\end{document}